\documentclass[universe,article,accept,moreauthors,10pt,a4paper]{mdpi} 
\firstpage{1} 
\articlenumber{x}
\doinum{10.3390/------}
\pubvolume{3}
\pubyear{2017}
\copyrightyear{2017}
\history{Received: 18 May 2017; Accepted: 3 July 2017; Published: date}

\usepackage{booktabs}
\usepackage{multirow}
\usepackage{soul}
\usepackage{microtype}

 \theoremstyle{mdpi}
 \newcounter{thm}
 \newcounter{ex}
 \newcounter{re}
\newcommand{\rmd}{\mathrm{d}}
\newcommand{\rmi}{\mathrm{i}}

\Title{Etherington's Distance Duality with Birefringence}
\Author{{Frederic P. Schuller $^{1,\dagger}$ and Marcus C. Werner $^{2,}$*$^{,\dagger}$} }
\AuthorNames{Frederic P. Schuller and Marcus C. Werner}
\address{%
$^{1}$ \quad Department of Physics, Friedrich-Alexander University Erlangen-Nuremberg, Staudtstrasse 7, 91058 Erlangen, Germany; frederic.p.schuller@fau.de \\
$^{2}$ \quad Yukawa Institute for Theoretical Physics and Hakubi Center for Advanced Research, Kyoto University, Kitashirakawa Oiwakecho Sakyoku, Kyoto 606-8502, Japan}

\corres{Correspondence: werner@yukawa.kyoto-u.ac.jp}
\firstnote{These authors contributed equally to this work.}

\abstract{We consider light propagation in a spacetime whose kinematics allow weak birefringence, and whose dynamics have recently been derived by gravitational closure. Revisiting the definitions of luminosity and angular diameter distances in this setting, we present a modification of the Etherington distance duality relation
in a weak gravitational field around a point mass. This provides the first concrete example of how the non-metricities implied by gravitational closure of birefringent electrodynamics affect observationally testable relations.}


\begin{document}




\section{Introduction}
Gravitational lensing and astrometry are closely related, since image positions, sizes and fluxes constitute some of the fundamental observables of gravitational lensing. Angular sizes and fluxes are connected with the corresponding lengths and luminosities of the observed object by angular diameter distances $D_A$ and luminosity distances $D_L$, respectively. Those distances, in turn, obey a well-known fundamental kinematical relationship with the redshift $z$, called the Etherington distance duality relation (e.g., \cite{sef99}, p. 116):

\begin{equation}
D_L = (1+z)^2 D_A\,,
\label{eth}
\end{equation}
a version of which was first derived by Etherington in 1933 \cite{e33}. This result is a purely kinematical one, since it depends only on Lorentzian spacetime geometry and a conservation law for light rays in geometrical optics, independent of the gravitational dynamics of general relativity. 

It has now become possible to perform observational tests of (\ref{eth}) in cosmology, for example by using the baryon acoustic feature as a standard ruler to measure $D_A$ and type Ia supernovae as standardizable candles to measure $D_L$ for given redshifts $z$. Such tests are particularly interesting since they provide a fairly direct probe of the fundamental kinematical structure of spacetime. Because of flux demagnification due to absorption, deviations from the standard Etherington relation towards lower $D_L$ are regularly expected. In addition, observations are also found to be consistent with deviations towards flux magnifications, and thus larger $D_L$, that may signal a fundamental spacetime property (e.g., \cite{mbh09}), although this effect appears not statistically significant at this point. In order to investigate this issue more deeply, it is necessary to study how such deviations might actually arise.

This issue has been considered, for instance, from the point of view of greybody spectra (e.g.,~\cite{epuw13}), and~modified gravity theories with scalar degrees of freedom (see, e.g.,~\cite{hp16} for observational and~\cite{bbdg13,mh14} for theoretical treatments).
Recently, we considered the most general electromagnetism that still satisfies a superposition principle and shows no birefringence. Contrary to previous results, we found, by careful derivation of the corresponding energy-momentum, that the Etherington relation remains unchanged in this case (\cite{mnssw17} and references therein), despite the presence of an axion and dilaton. 

In the present paper, we shall now consider a kinematical spacetime structure without the stipulation that birefringence be excluded. Such kinematics correspond to a premetric geometry or area metric geometry, which both refine the Lorentzian spacetime geometry of general relativity (e.g.,~\scalebox{.95}[1.0]{\cite{ho03,rrs11}}). 
Recent advances in the constructive gravity program  have shown how canonical gravitational dynamics for such refined backgrounds can be derived by the technique of gravitational closure of matter field equations on such a background (\cite{gsww12,sswd17} and references therein).

Specifically, we shall use the so derived weak-field dynamics for area metric spacetimes~\cite{sssw17}. The~pertinent gravitational background kinematics and the resulting  gravitational field around a point mass will be briefly described in Section~\ref{sec:background}. Following a discussion of how light propagates in this weak field limit, we derive the modified Etherington distance duality relation in Section~\ref{sec:duality}. We~use Latin indices to denote spacetime components and employ the metric signature $(+,-,-,-)$ when concerned with Lorentzian metrics.
\vspace{6pt}
\section{Gravitational Background}
\label{sec:background}
A consistent description of vacuum birefringence requires an understanding of the interplay between the pertinent electromagnetism and the dynamics of the refined spacetime geometry to which it couples. Fortunately, for quantizable matter dynamics, as we will consider, the need for shared initial data hypersurfaces and common evolution of both matter and geometry becomes so severely restrictive that the Lagrangian for the canonical dynamics of the geometry can be derived from the dynamics of the matter that lives on that geometry. This is the mechanism of gravitational closure~\cite{sswd17}. In~this~section, we~first formulate birefringent electrodynamics as the starting point for the central results of this paper and then report the form of the weak gravitational field around a point mass, as it follows from the perturbative solution the pertinent gravitational field equations obtained by gravitational closure.

\subsection{Kinematical Spacetime}
Consider a differentiable manifold, at first without any geometric structure in form of a metric or other, as a bare spacetime. Just as, in the standard theory, the Lorentzian spacetime geometry arises from Maxwell's electromagnetism in vacuo, we shall now see how a refined spacetime geometry arises from a generalized electromagnetism. The most general electromagnetism that still satisfies a superposition principle, and on such we wish to concentrate, has an action given by

\begin{equation}
S[A]=-\frac{1}{8}\int \rmd^4 x\ \omega_G G^{abcd}F_{ab}F_{cd},
\label{action}
\end{equation}
where the field strength tensor is $F_{ab}=2\partial_{[a}A_{b]}$ as usual, and $G$ is a tensor field of rank four with~symmetries

\[
G^{abcd}=-G^{bacd}, \qquad G^{abcd}=G^{cdab},
\]
which we shall regard as the refined geometrical structure of the underlying vacuum~spacetime. This~tensor field~has, in~fact, a~well-known geometrical interpretation as an (inverse) area~metric, by~which name it shall henceforth be referred~to, as~well as a physical interpretation in terms of a constitutive tensor describing effective material properties of an optical medium (e.g.,~\cite{p62}). This~area metric can also be used to define a volume form on our~spacetime, and~we shall use the~prescription 

\begin{equation}
\omega_G^{-1}=\frac{1}{24}\epsilon_{abcd}G^{abcd}
\label{volume}
\end{equation}
that has been employed in the above action. 

Light propagation in this theory is then governed by the principal polynomial tensor

\begin{equation}
P^{abcd}\propto \epsilon_{mn pq}\epsilon_{stuv} G^{mns(a} G^{b|pt|c} G^{d)quv},
\label{pptensor}
\end{equation}
also known as Fresnel or Tamm--Rubilar tensor (cf. \cite{ho03}, p. 267). Since the corresponding principal polynomial is quartic, light rays will exhibit birefringence in such a spacetime in general.

\subsection{Perturbative Dynamics}
We shall proceed by specializing the general area metric $G$ to a perturbation about a Minkowski metric background $\eta$, which may be written as

\begin{equation}
G^{abcd}=\eta^{ac}\eta^{bd}-\eta^{ad}\eta^{bc}-\sqrt{-\det \eta}\epsilon^{abcd}+ H^{abcd},
\label{linear}
\end{equation}
where $H$ is small. It turns out that it is useful to introduce the following quantities,

\[
\xi=\frac{1}{24}\epsilon_{abcd}H^{abcd}=\omega_G^{-1}-1, \qquad H^{ab}=H^{manb}\eta_{mn}, \qquad H=H^{ab}\eta_{ab}, 
\]
and~define

\[
h^{ab}=\frac{1}{2}H^{ab}-\xi \eta^{ab},
\]
which may be regarded as a metric perturbation on the Minkowski~background, giving~rise to the~fields 

\begin{equation}
P^{ab}=\eta^{ab}+h^{ab} \qquad\textrm{ and }\qquad P^\sharp_{ab}=\eta_{ab}-h_{ab}\,,
\label{p}
\end{equation}
which carry all zeroth and first order perturbation information of the kinematically relevant principal polynomial and dual polynomial

\[
P^{abcd}\propto P^{(ab}P^{cd)}+ \mathcal{O}(H^2) \qquad\textrm{ and }\qquad P^\sharp_{abcd}\propto P^\sharp_{(ab}P^\sharp_{cd)}+ \mathcal{O}(H^2) .
\]
The action for a light ray trajectory $x(\lambda)$ in the considered weakly birefringent background thus effectively becomes, to first order perturbation theory,

$$S[x,\mu] := \int d\lambda\, \mu\, P^\sharp_{ab} \dot x^a \dot x^b,$$
where $\mu(\lambda)$ is a Lagrange multiplier. This reveals $P^\sharp_{ab}$ as the effective metric geometry along whose geodesics light propagates.

Now, given this kinematical structure, the constructive gravity programme allows to explicitly derive the gravitational dynamics that underlie (\ref{action}), using the technique of gravitational closure of matter field equations with a bi-hyperbolic principal polynomial \cite{sswd17}. Employing the resulting gravitational action for a weak area metric derived in \cite{sssw17} and sourcing it by a point particle of mass $M$ at rest, one obtains \cite{amss17} the area metric refinement of the linearized Schwarzschild solution

\begin{eqnarray}
G^{0\alpha0\beta}&\equiv&-\gamma^{\alpha\beta}+H^{0\alpha0\beta} = -\gamma^{\alpha\beta}+(2A-\tfrac{1}{2} U + \tfrac{1}{2}V)\gamma^{\alpha\beta},  \label{g1} \\
G^{0\beta\gamma\delta}&\equiv& \epsilon^{\beta\gamma\delta}+H^{0\beta\gamma\delta} = -\epsilon^{\beta\gamma\delta}+\left(\tfrac{3}{4}U-\tfrac{3}{4}V-A\right)\epsilon^{\beta\gamma\delta}, \label{g2} \\
G^{\alpha\beta\gamma\delta}&\equiv& \gamma^{\alpha\gamma}\gamma^{\beta\delta}-\gamma^{\alpha\delta}\gamma^{\beta\gamma}+H^{\alpha\beta\gamma\delta} = (1+2U-V)(\gamma^{\alpha\gamma}\gamma^{\beta\delta}-\gamma^{\alpha\delta}\gamma^{\beta\gamma}), \label{g3}
\end{eqnarray}
where $\gamma^{\alpha\beta}$ and $\epsilon^{\alpha\beta\gamma}$ denote the Euclidean metric and Levi--Civita symbol of the~unperturbed, orientable~and flat three-dimensional spatial background, while the scalar perturbations

\begin{equation} 
A = - \frac{M}{4\pi r}\left(\kappa - \sigma \delta e^{-\mu r}\right)\,,\qquad
U = - \frac{M}{4\pi r} \delta e^{-\mu r}\,,\qquad 
V = \frac{M}{4\pi r}\left(\kappa - \tau\delta e^{-\mu r}\right)\label{scalarperturbations}
\end{equation}
in spherical polar coordinates $(r,\theta,\varphi)$. The constants $\sigma,\tau,\delta,\mu$ appearing in this solution arise as constants of integration in the gravitational closure procedure and would need to be determined~experimentally. The~final result of this paper provides, for instance, the theoretical underpinning for non-trivial experiments to this end. 
\vspace{6pt}
\section{Distance Duality}
Based on the kinematical and dynamical background outlined the previous section, we shall now proceed to derive the corresponding modification of the Etherington distance duality relation. Crucially, we shall need to consider the energy-momentum of light and its conservation law first. 
\label{sec:duality}
\subsection{Energy-Momentum}
The physical energy-momentum is technically encoded in a $(1,2)-$tensor density $\widetilde{T}^i{}_j$ and presents the reaction of the matter action to an infinitesimal diffeomorphism applied to the background~geometry. Thereby,~it~is directly related, but not identical, to the variation of the matter action with respect to the underlying geometry. For a thorough discussion, we refer the reader to \cite{gm,gm-2}. For our matter action (\ref{action}) on the area metric background $G$ in particular, the pertinent energy-momentum tensor density is given by 

\begin{equation}
\widetilde{T}^m{}_n=C^m{}_n{}^{abcd}\frac{\delta S}{\delta G^{abcd}},
\label{em}
\end{equation}
where $C$ is an intertwiner tensor that can be read off from the Lie derivative of the geometrical structure with respect to a shift vector $X$,

\[
(\mathcal{L}_X G)^{abcd}= X^n \partial_nG^{abcd}+ (\partial_m X^n) C^m{}_n{}^{abcd}.
\]
In the standard case of a metric geometry, one clearly recovers the standard definition of energy-momentum in terms of a variation of the action with respect to the metric using this prescription. However, in the case of our area metric structure, we obtain 

\begin{equation}
C^m{}_n{}^{abcd}=2\delta^{[a}_n G^{b]mcd}+2\delta^{[c}_n G^{d]mab}.
\label{intertwiner}
\end{equation}
Diffeomorphism invariance of the matter action then implies the conservation law 

\begin{equation}
\partial_m\widetilde{T}^m{}_n-\frac{\delta S}{\delta G^{abcd}}\partial_nG^{abcd}=0
\label{conservation}
\end{equation}
for the energy-momentum tensor density. Note that, due to the density weights, all equations are properly covariant despite being written in terms of partial derivatives. Indeed, there is no need to artificially construct a notion of a covariant derivative, neither in the standard theory nor~here, since~all equations follow without such theoretical~overhead.   

Having identified the appropriate notion of energy-momentum for our spacetime, we can now compute it from the action (\ref{em}). The corresponding equations of motion are a set of modified Maxwell's~equations,

\[
\partial_{b}\left(G^{abcd}F_{cd}\right)=0 \qquad\textrm{ and } \qquad \partial_{[b}F_{cd]}=0,
\]
and we can apply a WKB ansatz with eikonal phase function $S$ and $\epsilon\rightarrow 0$,

\[
F_{ab}=\mathrm{Re}\left[2k_{[a}A_{b]} \exp\left(\rmi \frac{S}{\epsilon} \right) \right],
\]
where $k_a=-\partial_a S$ is the wave covector as usual. Then, the energy-momentum can be found from (\ref{em}), and, upon averaging over the period of oscillation, one obtains

\begin{equation}
\langle \widetilde{T}\rangle^m{}_n=-\frac{1}{4}\omega_G G^{bmcd}(A_bA_c^\ast+A_b^\ast A_c)k_dk_n=:\widetilde{N}^mk_n,
\label{t}
\end{equation}
where the star denotes complex conjugation, and we can read off the photon current density~$\widetilde{N}^m$. Then,~the energy-momentum conservation law (\ref{conservation}) becomes

\begin{equation}
\widetilde{N}^m{}_{,m}=0.
\label{n1}
\end{equation}
Again, as the divergence of a vector density, this is covariant even though it is written in terms of a partial derivative. Coordinate-independence of this homogeneous equation is ensured by the tensor density property of $\widetilde{N}^m$. 

\subsection{Modified Etherington}
The properties of the photon current density defined above are crucial for the Etherington distance duality relation, which we shall now derive.

As a first step, we de-densitize the photon current density in order to obtain the photon current vector field

\[
N^m=\omega_G^{-1}\widetilde{N}^m,
\]
which is tangent to the bundle of light rays that are governed by the effective metric $P^\sharp_{ab}$ given by~(\ref{p}). 
Now, consider a bundle of light rays that congruently fills a spacetime domain $\mathcal{D}$ and is bounded by a timelike hypersurface $\partial \mathcal{D}_S$ containing a piece of the worldline of the light source $S$, where light rays enter $\mathcal{D}$, and similarly a timelike hypersurface  $\partial \mathcal{D}_O$ at the observer $O$, where light rays leave~$\mathcal{D}$. All~light rays emitted by the source during a certain proper time interval, and received by the observer during a corresponding interval, pierce these two hypersurfaces. 

Taking into account the non-trivial energy-momentum-stress transport encapsulated in (\ref{t}), one can then convert this trivial statement about geometric light rays into a non-trivial statement about photons, which ultimately yield the flux measured by the observer. Indeed, by comparing the integrated photon currents at $O$ and $S$, one obtains the excess fraction $\Delta$ of photons detected by the observer for any total number $N$ of photons emitted by the source, according to  

\[
\Delta \cdot N=\int_{\partial \mathcal{D}_O}\rmd^3x  \sqrt{\det p^\sharp}\  N^m n_m -\int_{\partial \mathcal{D}_S}\rmd^3x \sqrt{\det p^\sharp} \ N^m (-n_m),
\]
where  $p^\sharp$ is the metric of Euclidean signature induced from the effective spacetime metric $P^\sharp$ on the three-dimensional boundary hypersurfaces, and $n_m$ denotes the normal covector field. However, note that the part of the boundary of $\mathcal{D}$ other than $\partial \mathcal{D}_S, \ \partial \mathcal{D}_O$ is traced out by light rays and is hence null with respect to the effective spacetime metric $P^\sharp$, to the relevant order perturbation~theory. Therefore,~the~integration can be extended over the entire boundary $\partial \mathcal{D}$ whence

\begin{equation}
\Delta \cdot N = \int_{\partial \mathcal{D}}\rmd^3x \sqrt{\det p^\sharp} \ N^m n_m =\int_{\mathcal{D}} \rmd^4x \sqrt{-\det P^\sharp}\  N^m{}_{;m},
\label{delta}
\end{equation}
by applying Stokes' theorem (note the appearance of both $p^\sharp$ and $P^\sharp$), where the covariant differentiation is with respect to the light ray metric $P^\sharp$ in spacetime.

Now, given that the geometry of light rays is defined by $P^\sharp$, we can largely follow the standard derivation of the Etherington relation (cf. \cite{sef99}, p. 115f). Indeed, one can express the observed flux $S$ of a light source with luminosity $L$, which stems from the Poynting vector that is obtained from $\langle \widetilde{T} \rangle^0{}_\alpha$ in~(\ref{t}), as

\begin{equation}
S=\frac{(1+\Delta)L}{4\pi (1+z)^2 D^2},
\label{s1}
\end{equation}
where the distance $D$ is related to the angular diameter distance $D_A$ according to the standard~relationship,

\[
D=(1+z)D_A,
\]
since both are ray-geometrical quantities governed by the metric $P^\sharp$. On the other hand, one can define the luminosity distance $D_L$ in an arbitrary spacetime by

\begin{equation}
S=\frac{L}{4\pi D_L^2}.
\label{s2}
\end{equation}
Thus, combining (\ref{s1}) and (\ref{s2}), one obtains the Etherington distance duality relation

\begin{equation}
D_L=\frac{(1+z)^2D_A}{\sqrt{1+\Delta}},
\label{etherington1}
\end{equation}
where the excess fraction $\Delta$, which vanishes in the purely metric spacetime geometry of the standard~theory, remains~to be determined for our perturbative area metric spacetime geometry.

Specifically for a weakly birefringent gravitational field, one may concretely express the covariant divergence of the photon current vector in terms of the vector density and the gradient of a scalar perturbation of the background, namely as 

\begin{equation}
N^m{}_{;m} \,\,=\,\, \partial_m(\omega_G^{-1}\widetilde{N}^m)+\Gamma^n{}_{mn}\omega_G^{-1}\widetilde{N}^m \,\,=\,\, \partial_m \left(3\xi-\frac{H}{4}\right) N^m\,,
\label{n2}
\end{equation}
where the emerging connection coefficients are those of the Levi--Civita connection compatible with~$P^\sharp$. The~result (\ref{n2}) follows from the conservation law (\ref{n1}) and the weight of the volume form~(\ref{volume}). Even~more concretely, for the weakly birefringent gravitational field around a point mass $M$ given by Equations~(\ref{g1})--(\ref{g3}), we may recast the field

\[
3\xi-\frac{H}{4}\,\,=\,\,-3U\,\,=\,\, - \frac{3 M}{4\pi r} \delta e^{-\mu r},
\]
in terms of the Euclidean spatial distance $r$ from the point mass and the constants $\delta$ and $\mu$ only. Finally,~from~ (\ref{n2}) and (\ref{delta}), we can determine the excess fraction $\Delta$ in (\ref{etherington1}) and hence obtain

\begin{equation}
D_L=(1+z)^2D_A\left( 1+\frac{3\delta M}{8\pi}\left(\frac{e^{-\mu r_{ML}}}{r_{ML}}-\frac{e^{-\mu r_{MO}}}{r_{MO}}\right) \right),
\label{etherington2}
\end{equation}
as the Etherington distance duality relation modified by Yukawa terms, for the perturbative area metric spacetime of a point mass, where $r_{ML}$ is the Euclidean distance from the point mass to the light source and $r_{MO}$ from the point mass to the observer.
\vspace{6pt}
\section{Conclusions}
\label{sec:conclusion}
The most general electromagnetism that still satisfies a superposition principle, but admits vacuum birefringence, 
gives rise to unique kinematics and virtually unique dynamics of the underlying area metric spacetime geometry. The stationary gravitational field generated by a resting point mass, which we considered as a particular scenario only for definiteness, is determined up to only four constants, which would need to be determined by experiment, like both the cosmological and Newton constant need to be determined in the standard theory. 
By careful derivation of the energy-momentum tensor density and its transport, we computed a modified Etherington distance duality relation (\ref{etherington2}). Although, in the weak gravitational field regime considered here, the geodesics of light rays are still governed by an effective metric, we found that the area metric structure yields a non-metric correction to the Etherington relation. Our result provides the first example of a non-metric Etherington relation, whose derivation is based on a gravity theory that has been obtained from the pertinent matter dynamics by gravitational closure.

Since the Etherington relation is observable, it provides an important test of the fundamental spacetime structure. Of course, for this purpose, an area metric cosmological spacetime would be more interesting than the area metric point mass spacetime considered~here. We~anticipate that further results from the constructive gravity program, namely the exact solution of its central construction~equations, will~enable us to address this issue in the near future. On the other~hand, the~formalism for light deflection by a point mass in this weakly birefringent spacetime has implications for gravitational~lensing, which~we will address in a forthcoming article \cite{sw17}.

\vspace{6pt} 

\acknowledgments{F.P.S. thanks the Yukawa Institute for Theoretical Physics, Kyoto University, for their generous hospitality during a two-month research stay in the autumn of 2016, where a part of this work has also been done.}

\authorcontributions{The authors have contributed equally to this paper.}

\conflictsofinterest{The authors declare no conflict of interest.} 

\bibliographystyle{mdpi}

\renewcommand\bibname{References}

\end{document}